# Photometric Characterization and Trajectory Accuracy of Starlink Satellites: Implications for Ground-Based Astronomical Surveys

**Grace Halferty**
*Department of Aerospace and Mechanical Engineering, University of Arizona, Tucson, AZ, USA*
**Vishnu Reddy**
*Lunar and Planetary Laboratory, University of Arizona, Tucson, AZ, USA*
**Tanner Campbell**
*Department of Aerospace and Mechanical Engineering, Lunar and Planetary Laboratory,*
*University of Arizona, Tucson, AZ, USA*
**Adam Battle**
*Lunar and Planetary Laboratory, University of Arizona, Tucson, AZ, USA*
**Roberto Furfaro**
*Department of Systems and Industrial Engineering, Department of Aerospace and Mechanical*
*Engineering, University of Arizona, Tucson, AZ, USA*

## ABSTRACT

Starlink is a low-Earth orbit (LEO) satellite constellation operated by Space Exploration Technologies Corp. (SpaceX) which aims to provide global satellite internet access. Thus far, most photometric observations of Starlink satellites have primarily been from citizen scientists' visual observations without using quantitative detectors. This paper aims to characterize Starlink satellites and investigate the impact of mega constellations on ground-based astronomy, considering both the observed magnitude and two-line element (TLE) residuals. We collected 353 observations of 61 different Starlink satellites over a 16-month period and we found an average GAIA G magnitude of 5.5±0.13 with a standard deviation of 1.12. The average magnitude of V1.0 (pre-VisorSat) Starlinks was 5.1±0.13 with a standard deviation of 1.13. SpaceX briefly used a low-albedo coating on a Starlink satellite called DarkSat to test light pollution mitigation technologies. The brightness of DarkSat was found to be 7.3±0.13 with a standard deviation of 0.78, or 7.6 times fainter than V1.0 Starlinks. This concept was later abandoned due to thermal control issues and sun visors were used in future models called VisorSats. The brightness of VisorSats was found to be 6.0±0.13 with a standard deviation of 0.79, or 2.3 times fainter than V1.0 Starlinks. Over the span of the observations, we found that TLEs were accurate to within an average of 0.12 degrees in right ascension and –0.08 degrees in declination. The error is predominantly along-track, corresponding to a 0.3 second time error between the observed and TLE trajectories. Our observations show that a time difference of 0.3±0.28 seconds is viable for a proposed 10 second shutter closure time to avoid Starlinks in images.

# 1. INTRODUCTION

Starlinks are considered a mega constellation — a constellation composed of several hundreds to thousands of satellites orbiting the Earth, with a common mission. SpaceX began launching Starlink satellites in May 2019, with approval to launch up to 12,000 satellites (Mann, 2022). However, their hope is to eventually extend this to up to 42,000 satellites in low-Earth orbit (LEO). As of April 2022, over 2,335 Starlink satellites had been launched (McDowell, 2022). For comparison, there are >4,800 active satellites currently orbiting Earth, including the aforementioned operational Starlinks (Mann, 2022). The nominal schedule calls for two Starlink launches a month, each deploying between 49-60 satellites per launch (Grush, 2020). Since Starlink aims to provide internet service with the lowest latency possible, these objects need to orbit in LEO to reduce the information travel time. However, satellites in LEO can adversely impact ground-based astronomy as they are relatively bright and have high angular velocity. The deployment of such large satellite constellations has important implications for ground-based astronomical surveys as the bright Starlink satellites can produce image trails, ruining both imaging and spectroscopy data. (Mann, 2022).

Starlink satellites are expected to dominate orbits below 600 km, with a density of up to 1 satellite per 100 square degrees according to McDowell (2020). At low elevations near twilight observed from intermediate latitudes, hundreds of satellites may be visible at once even to the naked eye. The potential impact of Starlink satellites being in a low orbit (brightness and angular velocity) is compounded due to the sheer volume of satellites planned for launch. One large astronomical survey most at risk is the Legacy Survey of Space and Time (LSST) to be carried out by the Vera C. Rubin Observatory. LSST involves scanning of the entire night sky down to visual magnitude 25 in a single exposure every few days (Jones, 2017). This survey is noted as one of the many that could be negatively impacted by the increasing deployment of satellites in mega constellations. In the early days after the launch of the first Starlink satellites there was a lack of quantitative satellite brightness measurements to assess the actual effect they might have on ground-based astronomy (Grush, 2020). Most Starlink brightness measurements thus far have been qualitative, using just the naked eye, rather than using telescopes and CCD or CMOS cameras. There have been a few observational and model-based photometric studies by Tregloan-Reed, et al. (2020), Hainaut and Williams (2020), and Horiuchi, et al. (2020). In addition, there are several more from online archives that are not peer-reviewed, such as Mallama (2021), Krantz, et al. (2021), and Bassa, et al. (2021). Additionally, there have been no rigorous astrometric studies of Starlink satellites.

A possible mitigation strategy that has been proposed is to close the camera shutter on large survey telescopes when a Starlink passes through the field of view to prevent the satellites from saturating the sensitive detector (Tyson, et al., 2020). This requires precise knowledge of all Starlink satellites positions so the imaging can be paused at the appropriate times and avoid loss of survey time. Although the US Space Command and SpaceX track Starlink satellites and provide two-line elements (TLEs), their uncertainty is not provided, and could prove to be unreliable for pausing astronomical imaging at survey telescopes.

Tracking satellites in LEO is challenging since the objects move very quickly (approximately 200 arcminutes per minute) through a typical telescope's small field of view. In addition, it is challenging to extract accurate photometry from trailed images, as one has to decide between rate tracking on the satellite versus sidereal tracking, and at the same time avoiding detector saturation. Sidereal tracking keeps the star background stationary in images, which gives a good plate solution but results in the satellites appearing as streaks. Rate tracking requires prior orbital knowledge of the satellite, and results in trailed stars which again typically results in less accurate plate solution than using sidereal tracking (Campbell, 2018).

In response to concerns from the astronomical community, SpaceX has made some effort to mitigate the impact of their V1.0 satellites by attempting to reduce their brightness (throughout this study, "V1.0" Starlinks refer to pre-VisorSat and non-DarkSat satellites). One attempt was to coat one of the satellites, dubbed "DarkSat", with a low albedo material thereby hoping that it will reduce the satellite's reflectivity. The dark coating caused thermal control issues, however, and subsequently SpaceX implemented deployable sun visors instead. These satellites, known as "VisorSats", reduce brightness by blocking the sun glare reflected towards the Earth. All Starlinks launched between 2020 Aug. 7 and 2021 Jun. 30 are VisorSats (Young, 2020; Gohd, 2021). All of the observations reported in this paper are of Starlinks launched prior to the 2021 Jun. 30 switch in model.

The goal of our project is to answer the following four science questions. What is the apparent GAIA G magnitude of V.10 Starlink satellites? What is the apparent GAIA G magnitude of DarkSat and what does this imply about the effectiveness of the dark coating on Starlink satellites? What is the apparent GAIA G magnitude of VisorSat and how does this show the effectiveness of sun visors in reducing the satellite's apparent brightness? What is the accuracy of the published TLEs in predicting the location of Starlink satellites? We chose the GAIA G photometric catalog because it is the most modern and accurate catalog currently available.

## 2. HARDWARE AND OBSERVATIONS

Observations for this research were collected with the Stingray prototype located in Tucson, Arizona. This prototype consists of a 16-megapixel ASI 1600 MM Pro CMOS camera coupled with a Sigma 135 mm f/1.8 lens yielding a field of view of 7.5° x 5.7° and plate scale of 5.81"/pixel (Fig. 1). This shutterless, wide field of view sensor allows us to collect multiple 0.2-second sidereally-tracked images of each Starlink satellite pass without excessive trailing of the satellite while still collecting enough background stars for in-frame photometric calibration and astrometric plate solution. The filter used was a Sloan g' photometric filter. Dates of observations and targets observed can be found in Table 1. Observations were collected for almost every launch batch number up until June 2021. For each satellite, the latest TLE was pulled from Space-Track.org just before observations were taken, meaning the most up-to-date information available was used. These TLEs were then saved and used in the error analysis.

For planning observations, we checked for Starlink passes in one week intervals and looked for Starlinks 30 degrees above the horizon and in the western part of the sky since they would otherwise be shadowed in the eastern part. We observed when the Starlink was at its maximum altitude. There were 31 times where we attempted to observe a Starlink and either the object did not pass within our field-of-view or what was observed was not the targeted Starlink but some different object altogether. Orbit determination was used to rule out objects in the images we were confident were not Starlinks, i.e. moving in the wrong direction or with the wrong angular rate. The Stingray prototype's limiting magnitude is roughly 10 at 0.2 seconds exposure with SNR of ~15. Since the images were sidereally-tracked, the limiting magnitude for the trailed objects will be slightly brighter, however we expect that we would still be able to measure Starlinks as faint as magnitude 9 with a similar SNR. The object could also have an inaccurate TLE or could be fainter than our detection threshold. These missed objects are not considered or included in our analysis.

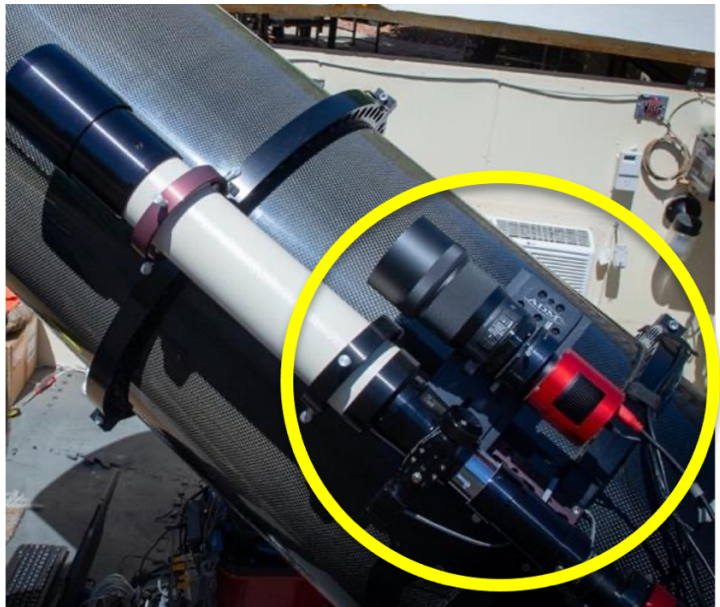

**Figure 1.** The Stingray prototype used to observe Starlinks.

**Table 1.** Observational circumstances for Starlink satellite passes.

| Date (UTC) | # of Satellites | Launch Batch #* | Days Since Launch |
|---|---|---|---|
| 2021 Feb. 6 | 4 | 1,2 | 453, 396 |
| 2021 Feb. 7 | 5 | 1,2 | 454, 397 |
| 2021 Feb. 8 | 4 | 3,8 | 376, 240 |
| 2021 Feb. 25 | 1 | 6 | 309 |
| 2021 Feb. 26 | 1 | 11 | 176 |
| 2021 Mar. 6 | 4 | 4,5,9 | 383, 353, 211 |
| 2021 Mar. 28 | 1 | 3 | 424 |
| 2021 Mar. 29 | 2 | 9,12 | 447, 174 |
| 2021 Mar. 11 | 11 | 0,3,15,18 | 677, 427, 126, 55 |
| 2021 Apr. 1 | 5 | 10 | 226 |
| 2021 Apr. 17 | 5 | 4 | 425 |
| 2021 Apr. 18 | 2 | 3 | 445 |
| 2021 Apr. 20 | 2 | 13 | 184 |
| 2021 May 7 | 5 | 2,9 | 486, 273 |
| 2021 May 8 | 5 | 9 | 274 |
| 2021 May 9 | 2 | 4 | 447 |
| 2021 Jun. 13 | 2 | 22 | 81 |

* Batch number refers to the Starlink V1.0 launch sequence, beginning with V1.0-L1 on 2019 Nov 11 (Krebs, 2022). A more extensive table can be found in Appendix A.

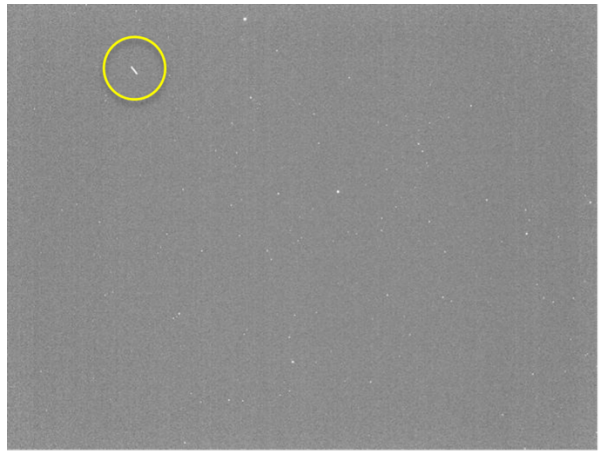

**Figure 2.** Example image of data collected for Starlink satellite STARLINK-1235 on 2021 Apr. 17.

An example of the type of image obtained with Stingray is shown in Figure 2. The data collection involved using a planetarium program called *TheSkyX* to check for satellite visibility and construct observation plans. In addition, we used the website *Heavens Above* that gave 10-day predictions of Starlink passes for a given location.

## 3. RESULTS AND METHODOLOGY

Apart from science images, dark and bias frames were also collected each night and flat field images were obtained once per month. Images were reduced following standard reduction techniques of bias and dark subtraction and flat-field division. Data analysis was carried out using a custom astrometric and photometric reduction pipeline (Campbell et al., 2018, 2019). The astrometric and photometric reduction was done using the GAIA DR2 catalog. For the photometric calibration, in-frame solar type stars were used based on the criteria published in Andrae, et al. (2018). A linear regression for each image set was used to calculate the zero point for correcting the measured instrumental magnitude to the GAIA G magnitude scale. A linear regression for each image set was used on the measured instrumental magnitude to correct to the GAIA G magnitude scale. The average of the RMS residual error for each set's linear regression is used as the representation of the photometric calibration of all observations. We find this systematic calibration uncertainty to be ±0.13 magnitude. Instrumental magnitude uncertainties were found to be negligible (i.e., < 0.002 mag) due to the brightness of the targets in the elliptical apertures. Circular apertures were used for the calibration stars, but an elliptical aperture was used for the streaked satellite trail. Source Extractor and Scamp were used as detailed in Campbell, et al. (2018, 2019).

### 3.1. PHOTOMETRIC STUDY

The photometric study involved using the images to extract the magnitude of the Starlink satellite. While the observations were made using a Sloan g' filter for observations, we were able to transform them to GAIA G magnitudes by using only solar-type stars and in-frame photometry. The following plots show data from 61 different Starlink satellites from 353 observations. Figure 3 represents average GAIA G magnitudes for each of the 61 Starlink satellites.

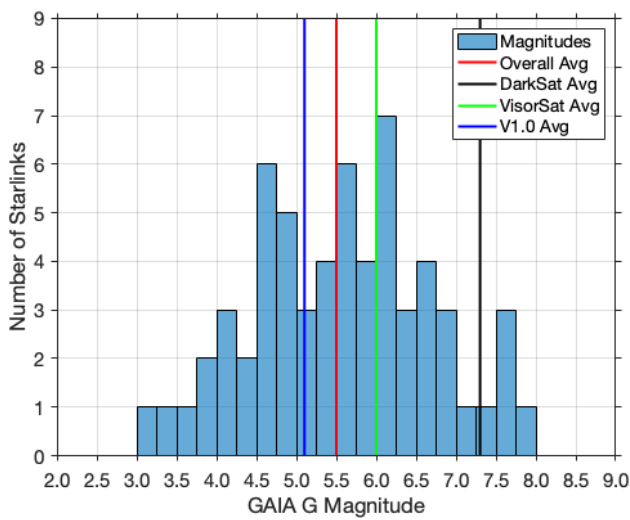

**Figure 3.** Histogram of averaged GAIA G magnitudes for 61 Starlink satellites. The red line represents the average magnitude for all observed Starlinks (5.5±0.13 with a standard deviation of 1.12). The blue line represents the average magnitude of V1.0 Starlinks, i.e., those that are not VisorSats or DarkSat (5.1±0.13 with a standard deviation of 1.13). The green line represents the average magnitude for VisorSats (6.0±0.13 with a standard deviation of 0.79), the black line represents the average magnitude for DarkSat (7.3±0.13 with a standard deviation of 0.78). It is important to note that since standard deviation is showing the variance in all measurements, it encompasses phase angle effects, atmospheric conditions, and true satellite-to-satellite variation.

Figure 3 shows the average GAIA G magnitudes for 61 Starlink satellites, with lines for average values for V1.0 Starlinks, VisorSat and DarkSat. It was found that the average magnitude for observed VisorSats was 6.0±0.13 with a standard deviation of 0.79. For DarkSat, the average magnitude was 7.3±0.13 with a standard deviation of 0.78.

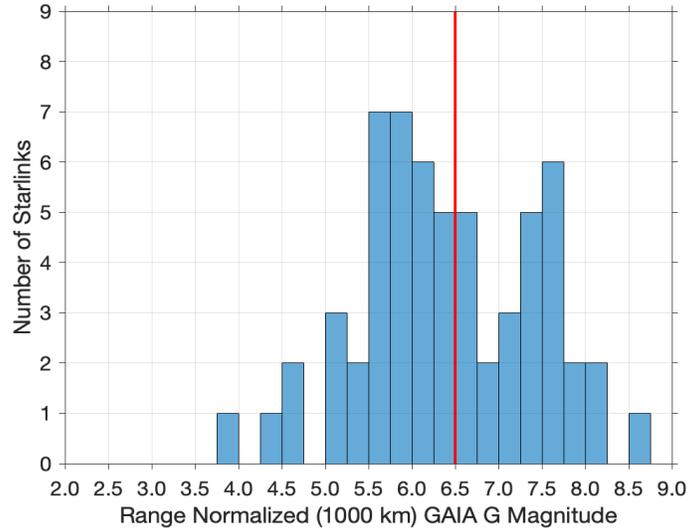

Figure 4. Histogram of range normalized (1000 km) GAIA G magnitudes for 61 Starlink satellites, with a line for the overall average (6.5±0.13 with a standard deviation of 1.40).

In completing the photometric study, we found that the average apparent GAIA G magnitude across all observed Starlinks was 5.5±0.13 with a standard deviation of 1.12. Additionally, the average magnitude of V1.0 Starlinks was 5.1±0.13 with a standard deviation of 1.13. The brightness of DarkSat was found to average 7.3±0.13 with a standard deviation of 0.78. This makes DarkSat 7.6 times fainter than V1.0 Starlinks. The brightness of VisorSat was found to be 6.0±0.13 with a standard deviation of 0.79. VisorSat is 2.3 times fainter than V1.0 Starlinks. Thus, the DarkSat design is a better mitigation method at reducing brightness. However, all Starlinks launched between 2020 Aug. 7 and 2021 Jun. 30, are VisorSats, due to thermal management issues in DarkSat (Mallama, 2021). The results show that although certain factors (such as sun visors on VisorSat) can reduce brightness, the satellites are still bright especially for large ground-based telescope observations.

Tyson, et al. (2020), reports the V1.0 Starlinks as g ~ 5.1 mag and DarkSat as g ~ 6.1 mag (using DECam, Victor M. Blanco 4-meter Telescope). Magnitudes brighter than V = 7 are predicted to have the most impact on observations (Lawler, et al., 2021). According to observations by Lawler, et al. (2021), over 70% of VisorSat Starlinks are brighter than V = 7, which is consistent with our study. Using Table 3 from Jordi, et al. (2010), a transformation for GAIA G magnitude to Johnson-Cousin V magnitude can be done, assuming solar color B-V = 0.653 for the satellites (Ramirez et al., 2012). The equation for this transformation is shown as Equation 1 with the transformation uncertainty reported by Jordi et al. (2010).

$$V = G + 0.235024 \ (\pm 0.38) \qquad (1)$$

Using this transformation, the average magnitude of V1.0 Starlinks is V = 5.3±1.2. Here, the reported uncertainty is from adding the uncertainty from the transformation (0.38), the photometric systematic calibration uncertainty (0.13), and the standard deviation of the measured population (1.13) in quadrature. Without any sort of darkening or visor treatment, V1.0 starlinks are still too bright by the limits proposed by Lawler et al. (2021) and continue to impact ground based observations. According to the conversions reported in Table 2, DarkSat would be the only suitable option to avoid the brightness level desired by astronomers, however this is not a viable option due to the aforementioned thermal management issues. Table 2 shows various reported Starlink magnitudes compared to what we have found.

**Table 2.** Reported V-magnitudes for V1.0 Starlinks, DarkSat, and VisorSat.

| Paper | V1.0 V-mag | DarkSat V-mag | VisorSat V-mag |
|---|---|---|---|
| This study | 5.3±1.2 | 7.5±0.88 | 6.2±0.89 |
| Krantz et al. 2021 | 7.0±1.1 | N/A | 8.0±1.1 |
| Tyson et al. 2020 | 5.3 (no reported uncertainty) | 6.3 (no reported uncertainty) | N/A |

Our average V magnitude of 5.3±1.2 is within 2σ uncertainty of Krantz, et al. 2021 and Tyson, et al. 2020. Any differences can potentially be explained by: phase angle effects, range differences, photometric processing, and color terms. Our study covers the phase angle range of -150 to 100 degrees, while Tyson, et al. (2020) uses 56 to 61 degrees, and Krantz, et al. (2021) uses ~30 to 170 degrees, resulting in different observing conditions that could affect the apparent magnitude. Since there was no common normalization of range among all of these study calculations, each study observes apparent magnitudes that can vary based on location and time of year. Streaked photometry is difficult and becomes harder to do with longer streaks. Shorter streaks, as used in this study, are closer to point source photometry and easier for accurate photometric extraction. In order to compare all three studies in Johnson's V-magnitude, a transformation was used which relies on a color estimate for the satellites which also leads to greater uncertainty.

### 3.2. ASTROMETRIC STUDY

For our astrometric analysis, images were plate solved using the reduction process outlines in Campbell, et al. (2018). The goal was to calculate along-track position error in degrees and seconds compared to the time after launch, as well as residual between published TLEs and measured positions.

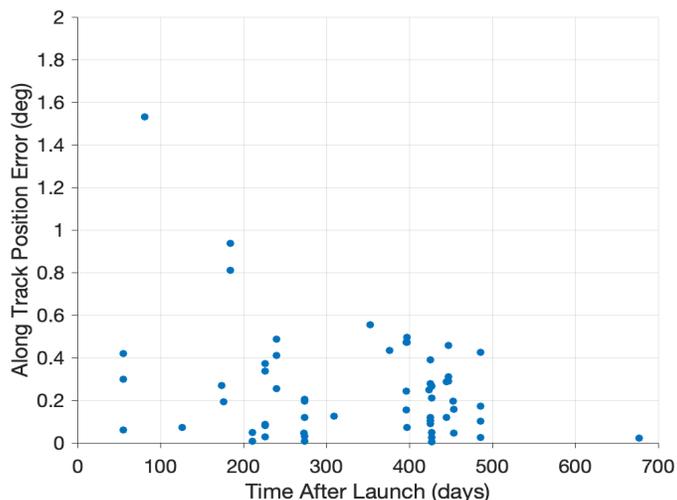

**Figure 5.** Along track deviation from the TLE in degrees versus time after launch. It is important to note that astrometric uncertainty (measured against TLEs) will be affected while Starlinks are still in their orbit raising phase.

Figure 5 shows the along-track deviation from the published TLE in degrees versus time after launch. From Figure 5, we can see there is not much of a trend between time after launch and position error for the observations we made. The time between launch date and observation seems to have little effect on positional uncertainty for the observed Starlinks assuming they have reached their final orbit. It is important to note that astrometric uncertainty (measured against TLEs) will be affected while Starlinks are still in their orbit raising phase. We also emphasize that positional uncertainties reported here could be affected by the TLE quality despite our best effort to use the most recent available prior to observations. Our finding has important implications because depending on when satellites are observed, and depending on how recent the TLE is during observations, the magnitude of this error could be vastly different.

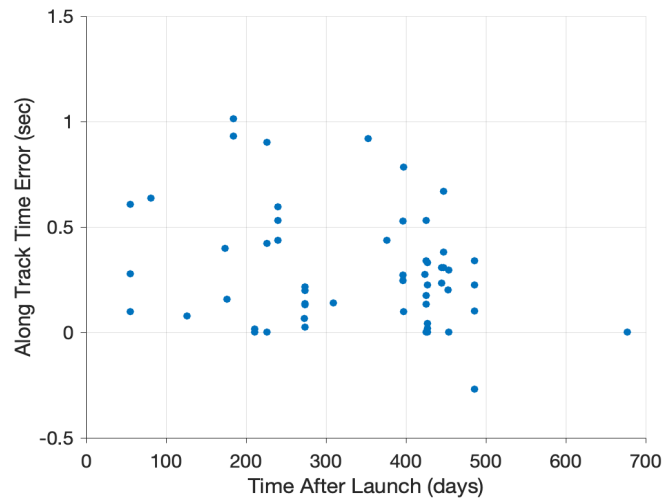

**Figure 6.** Along-track deviation from the TLE in seconds versus time after launch. It is important to note that astrometric uncertainty (measured against TLEs) will be affected while Starlinks are still in their orbit raising phase.

Figure 6 shows the along-track deviation from the published TLE in seconds versus time after launch. As in Figures 5 and 6, there is not much of a trend between time after launch and position error. The majority of our data is between 200-500 days after launch at which point it is most likely the Starlinks have reached their final orbit. Therefore these astrometric uncertainties are more representative of their final TLE uncertainties. Further information on the orbital life cycle can be found at Jonathan's Space Pages - see the table "List of all Starlink satellites and their orbital history"(McDowell, 2022). The reason for plotting the along-track error in seconds as well as degrees is because most of the error is along the object's path, meaning a time delay successfully captures the positional error much in the same way as comparing angles.

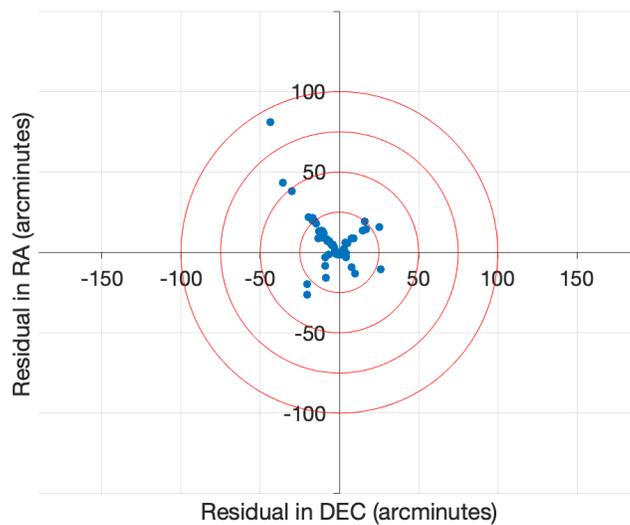

**Figure 7.** Bullseye plot of the deviation of the satellite position from the TLE in arcminutes. The concentric rings represent 25 arcminute increments.

Figure 7 shows the residual in right ascension (RA) in arcminutes versus the residual in declination (DEC) in arcminutes. This was done to see the deviation of the satellite position in both RA and DEC. Most Starlinks have a deviation within 25 arcminutes.

Following the astrometry analysis, the average difference in right ascension and declination between position measurements and the published TLE at epoch in degrees was found to be 0.12 and -0.08, with standard deviations of 0.26 and 0.20 degrees respectively, quantifying the difference between between published TLEs and the observed position. The average time difference between the published and observed position is 0.3 seconds, with a standard deviation of 0.28 seconds. The time difference is not unexpected because the BSTAR term in TLEs, which estimates the effect of atmospheric drag on a satellite, is difficult to quantify. The time between launch date and observation seems to have little effect on positional uncertainty during our observation campaign. This poses implications for observers trying to determine when to close telescope shutter for astronomical observations. These error results can be compared to a satellite-dodging scheme discussed in Tyson et al. (2020). This scheme details that the telescope scheduler for the Rubin Observatory pauses for 10 seconds for the Starlink to clear the field-of-view if a crossing will happen. Our observations show that a time difference of 0.3±0.28 seconds is well within the proposed 10 second shutter closure time. It is important to note that there could be longer than 0.3±0.28 seconds timing uncertainty with the Starlinks that were not detected, which could present issues for the proposed 10-second exposure pause solution described in Tyson et al. (2020). By quantifying the along-track time uncertainty of the Starlink TLEs, this study may help those planning the LSST and other surveys select a shorter closure window which would save valuable on-sky time for the surveys. A typical time difference of 0.3 seconds, with a 3 sigma high value of 1.2 seconds, should be a sufficiently small error that the observations will not be severely affected.

## 4. CONCLUSION

We collected 353 observations of 61 different Starlink satellites over a 16-month period. We found an average GAIA G magnitude of 5.5±0.13 with a standard deviation of 1.12. The average magnitude of V1.0 Starlinks was 5.1±0.13 with a standard deviation of 1.13. The brightness of DarkSat was found to average 7.3±0.13 with a standard deviation of 0.78, or 7.6 times fainter than V1.0 Starlinks. The brightness of VisorSat was found to be 6.0±0.13 with a standard deviation of 0.79, or 2.3 times fainter than V1.0 Starlinks. The average difference in right ascension and declination between position measurements and the published TLE at epoch in degrees was found to be 0.12 and -0.08, with standard deviations of 0.26 and 0.20 degrees respectively. The average time difference between the published and observed position is 0.3 seconds, with a standard deviation of 0.28 seconds.

The goal of this research was to characterize Starlink satellites and investigate the impact of mega constellations on ground-based astronomy. To characterize the satellites, both observed GAIA G magnitude and a comparison of reported versus measured positions were calculated. These results are important for large ground based astronomical surveys to consider because it can help them come up with mitigation strategies, such as closing camera shutters, to avoid Starlinks in their observations. The average time error is an important metric for beginning to understand how far the satellite's observed position deviates from the expected position. However, this metric likely only applies to those satellites that have reached their final orbit, where they are at their operational altitude. In addition, we found that the average differences in right ascension and declination measurements and the published TLE are 0.12 and -0.08 degrees.

Our study shows that a 10-second exposure pause to avoid Starlinks, as suggested by Tyson, et al. (2020) for Rubin Observatory, is a viable solution given the TLE along-track errors. While we only have a small sampling of each satellite's orbit, future work could focus on TLE uncertainty as a function of time from launch, since this is an important factor for newly launched satellites.

## ACKNOWLEDGMENTS

Thank you to the Air Force Research Lab (AFRL) Cooperative Agreement with the University of Arizona.

## DATA AVAILABILITY

The data underlying this article will be shared on reasonable request to the corresponding author.

# APPENDIX A

| Starlink Name | NORAD ID | Observation Date (UTC) | Launch Date | Days since Launch | Launch Date and Batch | Orbital Height (km) | Elevation (degrees) |
|---|---|---|---|---|---|---|---|
| SL-1032 | 44737 | 2/6/21 | 11/11/19 | 453 | Nov 11 2019, L1 | 547.219 | 79 |
| SL-1079 | 44937 | 2/6/21 | 1/7/20 | 396 | Jan 7 2020, L2 | 547.192 | 62 |
| SL-1094 | 44941 | 2/6/21 | 1/7/20 | 396 | Jan 7 2020, L2 | 547.239 | 48 |
| SL-1484 | 45756 | 2/6/21 | 1/7/20 | 396 | Jan 7 2020, L2 | 547.229 | 63 |
| SL-1047 | 44752 | 2/7/21 | 11/11/19 | 454 | Nov 11 2019, L1 | 547.252 | 41 |
| SL-1054 | 44759 | 2/7/21 | 11/11/19 | 454 | Nov 11 2019, L1 | 547.207 | 39 |
| SL-1073 | 44914 | 2/7/21 | 1/7/20 | 397 | Jan 7 2020, L2 | 547.249 | 63 |
| SL-1106 | 44923 | 2/7/21 | 1/7/20 | 397 | Jan 7 2020, L2 | 547.258 | 83 |
| SL-1130 | 44932 | 2/7/21 | 1/7/20 | 397 | Jan 7 2020, L2 | 547.125 | 40 |
| SL-1477 | 45754 | 2/8/21 | 1/29/20 | 376 | Jan 29 2020, L3 | 547.217 | 77 |
| SL-1493 | 45759 | 2/8/21 | 6/13/20 | 240 | June 13 2020, L8 | 547.239 | 82 |
| SL-1509 | 45766 | 2/8/21 | 6/13/20 | 240 | June 13 2020, L8 | 547.237 | 36 |
| SL-1521 | 45768 | 2/8/21 | 6/13/20 | 240 | June 13 2020, L8 | 547.261 | 88 |
| SL-1360 | 45588 | 2/25/21 | 4/22/20 | 309 | April 22 2020, L6 | 547.244 | 90 |
| SL-1763 | 46364 | 2/26/21 | 9/3/20 | 176 | Sep 3 2020, L11 | 403.660 | 90 |
| SL-1187 | 45219 | 3/6/21 | 2/17/20 | 383 | Feb 17 2020, L4 | 547.164 | 57 |
| SL-1255 | 45399 | 3/6/21 | 3/18/20 | 353 | March 18 2020, L5 | 547.208 | 51 |
| SL-1522 | 46027 | 3/6/21 | 8/7/20 | 211 | Aug 7 2020, L9 | 547.229 | 40 |
| SL-1581 | 46042 | 3/6/21 | 8/7/20 | 211 | Aug 7 2020, L9 | 547.221 | 36 |
| SL-1145 | 45066 | 3/28/21 | 1/29/20 | 424 | Jan 29 2020, L3 | 547.237 | 59 |
| SL-1526 | 46029 | 3/29/21 | 1/7/20 | 447 | Aug 7 2020, L9 | 547.225 | 60 |
| SL-1681 | 46559 | 3/29/21 | 10/6/20 | 174 | Oct 6 2020, L12 | 547.207 | 36 |
| SL-61 | 44249 | 3/31/21 | 5/24/19 | 677 | May 24 2019, L0 | 503.664 | 71 |
| SL-1133 | 45064 | 3/31/21 | 1/29/20 | 427 | Jan 29 2020, L3 | 547.226 | 33 |
| SL-1150 | 45067 | 3/31/21 | 1/29/20 | 427 | Jan 29 2020, L3 | 547.251 | 41 |
| SL-1161 | 45068 | 3/31/21 | 1/29/20 | 427 | Jan 29 2020, L3 | 547.249 | 60 |
| SL-1151 | 45081 | 3/31/21 | 1/29/20 | 427 | Jan 29 2020, L3 | 547.263 | 37 |
| SL-1927 | 46689 | 3/31/21 | 11/25/20 | 126 | Nov 25 2020, L15 | 547.193 | 37 |
| SL-1131 | 45047 | 3/31/21 | 1/29/20 | 427 | Jan 29 2020, L3 | 547.241 | 87 |
| SL-1169 | 45061 | 3/31/21 | 1/29/20 | 427 | Jan 29 2020, L3 | 547.233 | 59 |
| SL-1956 | 47557 | 3/31/21 | 2/4/21 | 55 | Feb 4 2021, L18 | 547.290 | 34 |
| SL-1995 | 47590 | 3/31/21 | 2/4/21 | 55 | Feb 4 2021, L18 | 547.155 | 30 |
| SL-2008 | 47603 | 3/31/21 | 2/4/21 | 55 | Feb 4 2021, L18 | 547.283 | 38 |
| SL-1605 | 46123 | 4/1/21 | 8/18/20 | 226 | Aug 18 2020, L10 | 547.238 | 60 |
| SL-1624 | 46130 | 4/1/21 | 8/18/20 | 226 | Aug 18 2020, L10 | 547.208 | 88 |
| SL-1637 | 46133 | 4/1/21 | 8/18/20 | 226 | Aug 18 2020, L10 | 547.254 | 54 |
| SL-1638 | 46134 | 4/1/21 | 8/18/20 | 226 | Aug 18 2020, L10 | 547.219 | 62 |
| SL-1596 | 46141 | 4/1/21 | 8/18/20 | 226 | Aug 18 2020, L10 | 547.218 | 68 |
| SL-1208 | 45205 | 4/17/21 | 2/17/20 | 425 | Feb 17 2020, L4 | 547.248 | 33 |
| SL-1215 | 45225 | 4/17/21 | 2/17/20 | 425 | Feb 17 2020, L4 | 547.242 | 58 |
| SL-1226 | 45229 | 4/17/21 | 2/17/20 | 425 | Feb 17 2020, L4 | 547.207 | 67 |
| SL-1227 | 45230 | 4/17/21 | 2/17/20 | 425 | Feb 17 2020, L4 | 547.192 | 76 |
| SL-1235 | 45232 | 4/17/21 | 2/17/20 | 425 | Feb 17 2020, L4 | 547.248 | 50 |
| SL-1176 | 45090 | 4/18/21 | 1/29/20 | 445 | Jan 29 2020, L3 | 547.209 | 89 |
| SL-1194 | 45101 | 4/18/21 | 1/29/20 | 445 | Jan 29 2020, L3 | 547.225 | 39 |
| SL-1781 | 46685 | 4/20/21 | 10/18/20 | 184 | Oct 18 2020, L13 | 547.266 | 90 |
| SL-1810 | 46709 | 4/20/21 | 10/18/20 | 184 | Oct 18 2020, L13 | 547.249 | 82 |
| SL-1097 | 44916 | 5/7/21 | 1/7/20 | 486 | Jan 7 2020, L2 | 547.211 | 31 |
| SL-1098 | 44917 | 5/7/21 | 1/7/20 | 486 | Jan 7 2020, L2 | 547.237 | 40 |
| SL-1123 | 44930 | 5/7/21 | 1/7/20 | 486 | Jan 7 2020, L2 | 547.215 | 35 |
| SL-1130 | 44932 | 5/7/21 | 1/7/20 | 486 | Jan 7 2020, L2 | 547.235 | 53 |
| SL-1554 | 46055 | 5/7/21 | 8/7/20 | 273 | Aug 7 2020, L9 | 547.251 | 53 |
| SL-1555 | 46032 | 5/8/21 | 8/7/20 | 274 | Aug 7 2020, L9 | 547.247 | 77 |
| SL-1582 | 46043 | 5/8/21 | 8/7/20 | 274 | Aug 7 2020, L9 | 547.202 | 67 |
| SL-1524 | 46047 | 5/8/21 | 8/7/20 | 274 | Aug 7 2020, L9 | 547.257 | 54 |
| SL-1574 | 46062 | 5/8/21 | 8/7/20 | 274 | Aug 7 2020, L9 | 547.271 | 42 |
| SL-1577 | 46063 | 5/8/21 | 8/7/20 | 274 | Aug 7 2020, L9 | 547.242 | 48 |
| SL-1234 | 45190 | 5/9/21 | 2/17/20 | 447 | Feb 17 2020, L4 | 547.249 | 69 |
| SL-1209 | 45206 | 5/9/21 | 2/17/20 | 447 | Feb 17 2020, L4 | 547.247 | 60 |
| SL-2283 | 48014 | 6/13/21 | 3/24/21 | 81 | March 24 2021, L22 | 349.847 | 77 |
| SL-2287 | 48035 | 6/13/21 | 3/24/21 | 81 | March 24 2021, L22 | 349.577 | 68 |

**Table 3.** Detailed observation log of every observed and processed target Starlink satellite.